\documentclass[11pt,twoside]{article}
\begin{document}\hbadness=10000\thispagestyle{empty}
\title{Note on the existence theorem in 
``Emergent Quantum Mechanics and Emergent Symmetries''}

\author{Hans-Thomas Elze \\ \ \\ 
        Dipartimento di Fisica, Universit\`a di Pisa \\
        Largo Pontecorvo 3, I-56127 Pisa, Italia          \\ \ \\ 
        {\it E-mail: elze@df.unipi.it}}

\date{}% It is always \today, today,

\maketitle

\begin{abstract} Recently 't\,Hooft  
demonstrated that ``{\it For any quantum system there 
exists at least one deterministic model that reproduces all its dynamics 
after prequantization''}. An extension is presented here which  
covers quantum systems that are characterized by 
a complete set of mutually commuting Hermitian operators ({\it beables}).  
We introduce the symmetry of beables: any complete set of beables is as 
good as any other one which is obtained through a real general linear group 
transformation. The quantum numbers of a 
specific set are related to symmetry breaking initial and boundary conditions 
in a deterministic model. The Hamiltonian, in particular, can be taken   
as the emergent beable which provides the best resolution of 
the evolution of the model universe.         
\end{abstract}

%%%%%%%%%%%%%%%%%%%%%%
\section{Introduction}
The existence theorem of 't\,Hooft concerns the Schr\"odinger equation 
for a quantum system with a $d$-dimensional Hilbert space: 
\begin{equation}\label{G1} 
\frac{\mbox{d}\psi}{\mbox{d}t}=-i\hat H\psi  
\;\;, \end{equation} 
where $\hat H$ denotes the Hamiltonian, a $d\times d$ 
matrix here. 

As demonstrated in Ref.\,\cite{EmergentQM},  
the dynamics of Eq.\,(\ref{G1}) is reproduced 
in a {\it deterministic system} with two degrees of freedom, one periodic variable, 
$\varphi\in [0,2\pi [$, and another real variable, $\omega$, which evolve 
according to the classical equations of motion:    
\begin{eqnarray}\label{G2}
&&\frac{\mbox{d}\varphi (t)}{\mbox{d}t}=\omega 
\;\;, \\ [1ex] \label{G3}
&&\frac{\mbox{d}\omega (t)}{\mbox{d}t}=-\kappa f(\omega )f'(\omega )\;\;,\;\;\; 
f(\omega ):=\det (\hat H-\omega )
\;\;, \end{eqnarray}  
with a parameter $\kappa >0$. 

The following argument is based on the observation that $\omega$ moves exponentially 
fast towards one of the eigenvalues of $\hat H$, since multiplying $f$ by minus one
times its derivative $f'$ makes all corresponding zeros 
attractive.\footnote{For an illustration of the 
behaviour of the right-hand side of Eq.\,(\ref{G3}), see Figure\,1 of 
Ref.\,\cite{EmergentQM}.} The initial condition for 
Eq.\,(\ref{G3}) determines which eigenvalue $E_i$ is approached, resulting    
in a limit cycle for $\varphi$ with period 
$T_i\equiv 2\pi\omega_i^{-1}=2\pi E_i^{-1}$. 

In order to proceed, two auxiliary operators are introduced: 
\begin{equation}\label{momenta} 
\hat p_\varphi :=-i\frac{\partial}{\partial\varphi}
\;\;,\;\;\;  
\hat p_\omega :=-i\frac{\partial}{\partial\omega}
\;\;, \end{equation} 
which do {\it not} correspond to classically observable quantities. 
We also define the operator $\hat h$: 
\begin{equation}\label{evolutionop} 
\hat h:=\omega\hat p_\varphi -\frac{\kappa}{2}
\{ f(\omega )f'(\omega ),\hat p_\omega \}
\;\;, \end{equation} 
with $\{ x,y\}:=xy+yx$. This operator generates the evolution described by the classical 
equations of motion (\ref{G2})--(\ref{G3}). Indeed, they can be written as: 
\begin{eqnarray}\label{G21}
&&\frac{\mbox{d}\varphi (t)}{\mbox{d}t}=-i[\varphi (t),\hat h] 
\;\;, \\ [1ex] \label{G31}
&&\frac{\mbox{d}\omega (t)}{\mbox{d}t}=-i[\omega (t),\hat h] 
\;\;, \end{eqnarray} 
with $[x,y]:=xy-yx$. Thus, the operator formalism, which is familiar in quantum theory,  
turns out to be useful in this classical context as well.  
Remarkably, the generator $\hat h$ is Hermitian, despite the  
dissipative character of the equations motion. 
  
It is natural to consider the Hilbert space on which these operators act and 
call its elements {\it prequantum states}. They can be employed as usual, 
if one wishes to calculate the observable properties of the classical system, 
which are functions ${\cal O}(\varphi ,\omega )$.   

Let us consider the evolution of those prequantum states $\psi$ which describe the 
trajectory of the classical system for an arbitrary but fixed initial condition:  
\begin{eqnarray}\label{prequ} 
\psi (\varphi ,\omega ;t)&=&
\sum_n e^{in\varphi}\psi_n (\omega ;t) 
\\ [1ex] \label{prequ1} 
&\stackrel{t\rightarrow\infty}{\longrightarrow}&
\sum_n e^{in(\varphi -\omega_it)}\psi_n (\omega_i;0)    
\;\;, \end{eqnarray} 
where $\omega_i$ is the particular fixed point to which $\omega (t)$ 
is attracted, depending on its initial condition;  
the Fourier transformation takes periodicity in the angular 
variable into account.  

Finally, we see that in a {\it superselection} sector, where the absolutely conserved 
``quantum number'' $n$ is fixed to a particular value $n'$, the prequantum states are  
directly related to the energy eigenstates of the quantum system described by Eq.\,(\ref{G1}): 
\begin{equation}\label{psiE} 
e^{-iE_it'}\psi (E_i )
=e^{-in'\omega_it}\psi_{n'}(\omega_i;0)    
\;\;, \end{equation}  
evolving in the usual way, with $t':=n't$.
Probabilistic superpositions of such prequantum states with different $\omega_i$ result 
in (mixed) quantum states showing interference.  

In conclusion, characteristic features of quantum systems described by 
Eq.\,(\ref{G1}) emerge here from the dissipative evolution of deterministic systems beneath. -- 
This completes our review of the existence theorem 
of Ref.\,\cite{EmergentQM}.   

Our extension of the existence theorem is partly motivated by the fact that the 
appearance of the set of eigenvalues $\{ E_i\}$ of the 
Hamiltonian in $\det (\hat H-\omega )=\prod_i(E_i-\omega )$ is puzzling, in the 
classical context of Eqs.\,(\ref{G2})--(\ref{G3}): While the existence theorem provides 
a deterministic dynamics that gives rise to the quantum mechanical one, it does   
give neither a (classical or other) interpretation of this set of numbers nor a mechanism 
which determines them at the pre-quantum level. -- Furthermore, the states of generic 
quantum mechanical objects are fully specified only by a set of quantum numbers which corresponds to a complete set of mutually commuting Hermitian operators. 
The present note addresses these aspects. 

For additional results and discussions concerning emergent quantum mechanics, 
we refer the reader to Refs.\,\cite{tHooft06}--\cite{tHooft88}, 
with further references therein. 

%%%%%%%%%%%%%%%%%%%%%%%%%%%%%%%%%
\section{The symmetry of beables}  
A complete characterization of the state of a quantum mechanical object 
with a finite number of degrees of freedom, generally, 
requires a set of simultaneous eigenvalues of a number of linearly independent and mutually 
commuting Hermitian operators, $\hat A_n,\;n=1,\dots ,N$, which are called {\it beables} 
\cite{BellBook}. 
These operators are here assumed to act on a finite dimensional Hilbert space.  

A specific set of beables, represented by $\vec A:=(\hat A_1,\dots ,\hat A_N)^t$, 
can be interpreted as the set of operator valued coordinates of a point in an 
abstract $N$-dimensional vector space. The intrinsic physical properties of the 
quantum mechanical object correspond to this particular point, independently 
of the chosen coordinates. 
 
%For example, take the operators representing the coordinates of the hydrogen  
%atom and its electron. Instead, it is advantageous to employ relative and center-of-mass 
%coordinates, if there are no external forces. 
Objects carrying spin and isospin or spin and orbital angular momentum provide examples with finite dimensional Hilbert spaces. Various choices of beables are possible, related to 
various ``coupling schemes'' which are well known in atomic and nuclear physics.  

Clearly, the selection of sets of beables is largely conditioned 
by empirical input, such as the observation of dynamical symmetries,   
and the attempt to obtain a concise mathematical theory of 
the objects under study. 

However, concerning foundational issues of quantum theory, 
it may be useful not to limit the attention to special choices of beables. Instead,   
we wish to treat all possible choices on an equal footing,  
implementing the {\it Symmetry of Beables}: 
\begin{itemize} 
\item {\it There is no absolute meaning attached to a given set of beables}, 
represented by $\vec A$, as compared to a second  
one, $\vec A'$, which is obtained from the former by the action of a group of symmetry 
transformations. The transformed set describes the quantum mechanical 
object just as well, in principle, provided the map is compatible with the Hilbert space 
properties of hermiticity, pairwise commutativity, and completeness of the 
beables \cite{BellBook}.       
\end{itemize}

\noindent
Since operators with zero eigenvalues are not excluded, some of the $\hat A$'s may not 
have an inverse. This limits the set of real transformation functions representing the symmetry 
of beables to those that can be expanded into series of multinomials in $\hat A$'s. While 
nonlinear transformations are possible, for simplicity, we restrict ourselves 
to the group of linear transformations: 
\begin{equation}\label{transf} 
\hat A_n'=\sum_{m=1}^NM_{nm}\hat A_m\;\;,\;\;\; M\in\mbox{GL}(N,\mathbf{R}) 
\;\;, \end{equation} 
i.e., which can be represented by regular real $N\times N$ matrices. 
 
We note in passing that the general linear group in 
$N$ dimensions contains the group of permutations for sets of $N$ elements as a 
subgroup. The latter plays the role of the diffeomorphism group in the context of 
causal sets \cite{Sorkin,SorkinRideout}. 
Therefore, we speculate that there might be a connection between 
the symmetry of beables which are attributed to quantum mechanical objects,  
especially atomistic ``events'', and the correspondent 
of general coordinate invariance for a fundamentally discrete spacetime. 
  
The quantum states of the object evolve 
according to the Schr\"odinger equation (\ref{G1}), where  
the Hamiltonian presently acts on the finite dimensional Hilbert space 
pertaining to the set of beables. Given the symmetry, cf. Eq.\,(\ref{transf}), the 
Hamiltonian must be among the beables or be a linear combination of them. Otherwise 
they would not be complete.  

In the following sections, we will show how the theory of such quantum 
mechanical objects can be completely reconstructed in classical terms. 
In particular, the Schr\"odinger evolution and symmetry properties are shown to emerge 
from a deterministic prequantum model. Particular examples of emergent quantum 
mechanical symmetries have also been discussed in Refs.\,\cite{EmergentQM,Liu01}.    

%%%%%%%%%%%%%%%%%%%%%%%%%%%%%
\section{Prequantum dynamics} 
The model comprises $N$ real degrees of freedom which are periodic, 
$\vec\varphi :=(\varphi_1,\dots ,\varphi_N)^t$, $\varphi_n\in [0,2\pi [$, and 
evolve according to the classical equation of motion: 
\begin{equation}\label{phivec} 
\frac{\mbox{d}\vec\varphi (t)}{\mbox{d}t}=\vec\omega 
\;\;, \end{equation} 
involving a second set of $N$ real degrees of freedom, 
$\vec\omega :=(\omega_1,\dots ,\omega_N)^t$. The equation of motion of $\vec\omega$ 
will follow in Eq.\,(\ref{Fbc}) below. 
  
First, however, we introduce an auxiliary real field $F$ on the space of matrices $M$ representing 
the symmetry group of beables, $M\in\mbox{GL}(N,\mathbf{R})$, which obeys the     
wave equation: 
\begin{equation}\label{FwaveEq}  
\left (\partial_t+\dot{\vec\omega}\cdot\partial_{M\cdot\vec A}\right )
\left (\partial_t-\dot{\vec\omega}\cdot\partial_{M\cdot\vec A}\right )F(M,t)=0
\;\;, \end{equation}
where $\dot{\vec\omega}:=\mbox{d}\vec\omega /\mbox{d}t$. Parametrically, the  
field depends on real numbers $A_n^j$, $n=1,\dots,N$, $j=1,\dots,d$, 
which can be considered to define the sets of simultaneous eigenvalues of $N$ commuting 
Hermitian operators, collectively denoted by $\vec A:=(\hat A_1,\dots ,\hat A_N)^t$. 
This is also visible in the general solution of the wave equation: 
\begin{equation}\label{FwaveSol} 
F(M,t)=f(\vec\omega (t)-M\cdot\vec A)+g(\vec\omega (t)+M\cdot\vec A)
\;\;, \end{equation} 
where $f$ and $g$ are two real functions (of respective combinations  
of Hilbert space operators) with suitable differentiability properties, 
but otherwise arbitrary.  

An important property of beables is that related eigenvalues 
are invariant under unitary transformations in Hilbert space. With hindsight, 
therefore, we require the field $F$ to be a scalar under such transformations.  

Furthermore, the arguments $\vec\omega (t)\pm M\cdot\vec A$, which appear in 
the general solution (\ref{FwaveSol}), transform covariantly 
with respect to the symmetry of beables, i.e., under the transformations: 
\begin{eqnarray}\label{symm}
\vec A&\longrightarrow&\vec A'=S\vec A\;\;, \nonumber \\ 
\vec\omega&\longrightarrow&\vec\omega'=S\vec\omega\;\;, \nonumber \\ 
M&\longrightarrow&M'=SMS^{-1}\;\;,\;\;\;S\in\mbox{GL}(N,\mathbf{R})
\;\;, \end{eqnarray} 
leaving, of course, a suitably defined scalar product of vectors invariant. 

We now seek an {\it initial condition} that breaks the $\mbox{GL}(N,\mathbf{R})$ 
symmetry. This is motivated by the aim to eventually describe 
a quantum mechanical object with a fixed set of beables (related specifically to $\vec A$) 
in the deterministic model. Therefore, we would like 
the solution $F$ to be left invariant only by the subgroup of $\mbox{GL}(N,\mathbf{R})$ 
which effects permutations of the $\hat A_n$'s.  
%or 
%individual overall rescalings of them (represented by diagonal matrices). 
Yet, it must remain invariant under unitary transformations 
in Hilbert space. This leads us to the initial condition:      
\begin{eqnarray}\label{Finit} 
F(M,t_0)&\equiv&f(\vec\omega (t_0)-M\cdot\vec A)
\\ [1ex] \label{Finit1}
&\equiv&\sum_{n=1}^N\prod_{j=1}^d
[\sum_{m=1}^NM_{nm}A_m^j-\omega_n(t_0)]^2
\;=\;\sum_{n=1}^N{\det}^2[M\cdot\vec A-\vec\omega (t_0)]_n
\;\;, \end{eqnarray}   
where the last sum is over the components of the vector inside $[\dots ]$ and 
the determinant refers to a $d$-dimensional Hilbert space on which the  
operators act that are collected in $\vec A$. -- The evolving solution is then 
simply obtained by replacing $t_0$ by $t$ here, with $\vec\omega (t)$ 
still to be determined.   

It is important to realize that operators and Hilbert space have only 
been introduced for convenient bookkeeping.   
Essentially needed, so far, are arbitrary real numbers $A_n^j$ which 
parametrize the initial condition, as in Eqs.\,(\ref{Finit})--(\ref{Finit1}). 
  
Furthermore, we remark that the sum of squares of determinants in Eq.\,(\ref{Finit1})  
is zero, if and only if the $N$-dimensional vector $\vec\omega$ corresponds to  
one of the points of the $N$-dimensional finite lattice defined by the 
$d\times N$ numbers $A_n^j$ or, rather, by the $d$ eigenvalues of each one of the $N$ operators 
$M\cdot\hat A$. 
In this way, the initial condition here presents a generalization of the 
function $f(\omega)$ of 't\,Hooft's existence theorem, cf. Eqs.\,(\ref{G3}). 

Finally, a {\it boundary condition} on the solution 
for the field $F$ is imposed:  
\begin{equation}\label{Fbc} 
\frac{\mbox{d}\vec\omega (t)}{\mbox{d}t}=-\kappa\frac{\partial}{\partial\vec\omega}
F^2(M^\ast,t)
\;\;, \end{equation} 
with $\kappa >0$, and where $M^\ast\in\mbox{GL}(N,\mathbf{R})$ is fixed but arbitrary.  
This equation simultaneously determines $\vec\omega$, once    
its initial value $\vec\omega (t_0)$ is supplied.  
  
For the solution specified by the above initial condition, Eqs.\,(\ref{Finit})--(\ref{Finit1}), 
this boundary condition suitably generalizes Eq.\,(\ref{G3}). Similarly to our discussion in 
the Introduction, 
it is easy to see that the zeros of $F$ (corresponding to points on the  
$d\times N$ lattice above) are attractive. Thus, 
the vector $\vec\omega$ is attracted to a fixed point: 
\begin{equation}\label{fixp} 
\omega_n (t)\;\stackrel{t\rightarrow\infty}{\longrightarrow}\; 
\sum_{m=1}^NM^\ast_{nm}A_m^{j(m)}=:\omega_n^\ast 
\;\;, \end{equation} 
i.e., to a fixed vector with components built from linear combinations 
of eigenvalues of the operators $\hat A_m$; which particular eigenvalues 
contribute, indexed by $j(m),\;j=1,\dots,d$, depends on the arbitrary initial condition 
for $\vec\omega$, analogous to the case reviewed in the Introduction.     

Furthermore, the field $F$ decays to zero on the boundary ($M=M^\ast$) and approaches  
a constant value, $F(M,t\rightarrow\infty )=f(\vec\omega^\ast -M\cdot\vec A)$, elsewhere.  
Through the zeros of $F$ and depending on $\omega (t_0)$, 
asymptotically $\vec\omega\approx\vec\omega^\ast$ also defines a vector of 
eigenvalues for the $N$ operators $M\cdot\vec{A}$, simply given by 
$M\cdot (M^\ast )^{-1}\vec\omega^\ast$, 
for any choice of $M\in\mbox{GL}(N,\mathbf{R})$. 

This completes the set-up of the deterministic model. Before we turn 
to the emergent quantum mechanical behaviour in the next section,   
let us briefly discuss here the role of the auxiliary field $F$. 

Applying {\it Ockham's razor}, we should omit wave equation (\ref{FwaveEq}), since 
only Eqs.\,(\ref{phivec}), (\ref{Finit})--(\ref{Finit1}), and (\ref{Fbc}) are needed for what   
follows. However, only a field equation of motion allows to separately interpret 
Eqs.\,(\ref{Finit})--(\ref{Finit1}) and 
(\ref{Fbc}) as a symmetry breaking initial condition and a boundary condition, respectively. 
The boundary condition introduces dissipation which leads to the 
attractive periodic orbits of $\vec\varphi$ that will be essential for the 
quantum mechanical features. This may open a way to 
explain the dissipation as a dynamical effect of neglected degrees of freedom or 
nonlinearities that must come into play when one tries to deal with physical forces. 
The latter are still missing to some extent. Despite that all spectral information 
seems to arise from the initial condition (in the form of numbers $A_n^j$), 
their allowed values should be dynamically constrained, as well as the special  
symmetry breaking.\footnote{Unless we are prepared to accept an 
accidental yet decisive influence of the initial conditions on the model universe and 
its emergent (quantum mechanical) laws. The view that a random Hamiltonian might serve 
as a starting point to explain emergent physical laws has recently been expressed   
in Ref.\,\cite{Albrecht07} (see also further references therein), 
studying consequences of the clock ambiguity in time reparametrization invariant 
theories.} 

%%%%%%%%%%%%%%%%%%%%%%%%%%%%%%%%%%%%
\section{Emergent quantum mechanics}
The considerations of Eqs.\,(\ref{prequ})--(\ref{psiE}) are now easily generalized 
for the case at hand. 

Similarly as before, we consider prequantum states $\psi$ which describe the trajectory 
of the deterministic system for an arbitrary but fixed initial condition: 
\begin{eqnarray}\label{prequF} 
\psi (\vec\varphi ,\vec\omega ;t)&=&
\sum_{\vec n} e^{i\vec n\cdot\vec\varphi}\psi_{\vec n} (\vec\omega ;t) 
\\ [1ex] \label{prequF1} 
&\stackrel{t\rightarrow\infty}{\longrightarrow}&
\sum_{\vec n} e^{i\vec n\cdot(\vec\varphi -\vec\omega^\ast t)}\psi_{\vec n} 
(\vec\omega^\ast;0)    
\;\;, \end{eqnarray} 
where $\vec\omega^\ast$ is the fixed point to which $\vec\omega (t)$ 
is attracted, depending on its initial value $\vec\omega (t_0)$, as discussed in 
the previous section, and periodicity in $\vec\varphi$ underlies the Fourier transformation.  
  
A new feature arises here in that the states fall into  
superselection sectors that can be classified by the absolutely conserved vector $\vec n$. 
Furthermore, the states are specified by the asymptotic vector $\vec\omega^\ast$. 
All its components contribute 
to the phase, which describes the evolution of the state. That is, all $N$ Hermitian 
operators $\sum_mM_{nm}\hat A_m$ contribute, each with a set of $d$ eigenvalues 
$\sum_mM_{nm}A_m^j$. Following a single realization of the deterministic trajectory,  
the resulting particular eigenvalue can be given by 
$\sum_{m,k}M_{nm}(M^\ast )^{-1}_{mk}\omega^\ast_k$, as before.    

Three qualitatively different situations may arise. -- First, the model universe may 
find itself in a state where all components of $\vec n$ are equal, denoted by 
$\vec n'\equiv (n',\dots,n')^t$. In this case, the 
Hamiltonian, which generates the evolution, must be identified as: 
\begin{equation}\label{Hamiltonian} 
\hat H = \sum_{m,k=1}^NM_{mk}^\ast\hat A_k 
\;\;, \end{equation}  
which picks an eigenvalue $E_*=\sum_{m,k}M_{mk}^\ast A_k^{j(k)}=\sum_m\omega_m^\ast$, 
following a particular deterministic trajectory. 
Here, emergent quantum states are related to the prequantum states by: 
\begin{equation}\label{psiE1}
e^{-iE_*t'}\psi (E_*)=e^{-in'\sum_m\omega_m^\ast t}
\psi_{\vec n'}(\vec\omega^\ast ;0)     
\;\;, \end{equation} 
with $t':=n't$, cf. Eq.\,(\ref{psiE}). One of the beables, corresponding 
to $\omega^\ast_1$, for example, could be eliminated in favour of the Hamiltonian 
and $E_*$, respectively, such that the above relation becomes 
$\psi (E_*)\propto \psi_{\vec n'}(E_*,\omega^\ast_2,\dots,\omega^\ast_N;0)$. 
Thus, we find degenerate energy eigenstates, which are further resolved by the 
eigenvalues of the $N-1$ remaining beables, i.e., by the values of 
$\omega^\ast_2,\dots,\omega^\ast_N$.         

Second, assuming that the natural scales of all $\omega^\ast_n$ 
are of the same order of magnitude, one of the components of the superselection 
vector $\vec n$, say $n_1$, may be very much larger than all others. In this case, 
it seems natural to consider the Hamiltonian: 
\begin{equation}\label{Hamiltonian1} 
\hat H_\sim =\sum_{k=1}^NM_{1k}^\ast\hat A_k 
\;\;, \end{equation}   
with eigenvalue(s) $E_*=\omega_1^\ast$, which presents a valid approximation, 
as long as only sufficiently small eigenvalues $\omega^\ast_{m>1}$ have to be   
taken into account, $|n_1\omega^\ast_1|>>\sum_{m=2}^N|n_m\omega^\ast_m|$. 
In this case:
\begin{equation}\label{psiE2} 
e^{-iE_*t'}\psi (E_*)=e^{-in_1\omega_1^\ast t} 
\psi_{\vec n}(E_*,\omega^\ast_2,\dots,\omega^\ast_N;0) 
\;\;, \end{equation}  
with $t':=n_1t$. 
That is, the contribution to the phase which is most sensitive to the 
evolution dominates all others. This leads again to degenerate energy eigenstates, 
which are resolved by the remaining beables.  

There will be 
only accidental degeneracies, if any, of the emergent energy eigenstates 
in the third case, when {\it all beables} possibly contribute to the Hamiltonian:    
\begin{equation}\label{Hamiltonian2} 
\hat H_{all} = \vec n\cdot M^\ast\cdot\vec A 
\;\;, \end{equation}  
with eigenvalues of the form $E_*=\vec n\cdot\vec\omega^\ast$. Here, we 
obtain: 
\begin{equation}\label{psiE3} 
e^{-iE_*t}\psi (E_*)=e^{-i\vec n\cdot\vec\omega^\ast t} 
\psi_{\vec n}(\vec\omega^\ast ;0) 
\;\;, \end{equation}  
where an arbitrary $\omega^\ast_i$ could be replaced by 
$E_*-\sum_{m\neq i}n_m\omega^\ast_m/n_i$, 
provided $n_i\neq 0$. Thus, in this most general case, there still exist a 
unique Hamiltonian and a related energy variable, 
which govern the evolution of the emergent states. However,  
all beables contribute in a simple but nontrivial way to these quantities.  

%%%%%%%%%%%%%%%%%%%%
\section{Conclusion}   
We have shown that 't\,Hooft's theorem -- stating the existence of a deterministic 
model that reproduces the dynamics of a given quantum system -- can be extended 
to cover objects that are characterized by sets of beables. The symmetry of beables, which 
we introduced, has been useful in formulating a correspondingly extended deterministic 
prequantum model.   

As we discussed, the deterministic model building appears at a sort  
of kinematical stage, so far, since the initial and boundary data that we invoked 
ask for an explanation. Possibly, the large symmetry of beables will be 
an ingredient for a theory that explains the necessary dissipation 
mechanism and constrains the initial data -- and, thus, the relevant beables -- 
in a dynamical way.      

%%%%%%%%%%%%%%%%%%%%%%%%%%%
 
%%%%%%%%%%%%%%

\end{document}